\documentclass[twocolumn,superscriptaddress,floatfix, showpacs, amsmath, amssymb, aps, prl]{revtex4-2}
\usepackage[backref=none,bookmarksnumbered=true,bookmarks=true,  	bookmarksopen=true,colorlinks=true,citecolor=blue,linkcolor=blue, 	anchorcolor=green,urlcolor=blue,unicode=false]{hyperref}

\usepackage{graphicx}
\usepackage{float}
\usepackage{dcolumn}
\usepackage{bm}
\usepackage{chemformula}
\usepackage{siunitx}
\usepackage{verbatim}
\usepackage[T1]{fontenc}
\usepackage[utf8]{inputenc}

\usepackage{times}

\begin{document}
\newcommand{\equalcontribution}{\textsuperscript{§}}
\author{Vibhuti N. Rai\equalcontribution}
\email{vibhuti.rai@fu-berlin.de}
\affiliation{Fachbereich Physik and Halle--Berlin--Regensburg Cluster of Excellence CCE, Freie Universit\"at Berlin, Arnimallee 14, 14195 Berlin, Germany}
\author{Junyoung Sim\equalcontribution}
\affiliation{Fachbereich Physik and Halle--Berlin--Regensburg Cluster of Excellence CCE, Freie Universit\"at Berlin, Arnimallee 14, 14195 Berlin, Germany}
\author{Nils Bogdanoff}
\affiliation{Fachbereich Physik, Freie Universität Berlin, 14195 Berlin, Germany}
\author{Sergey Trishin}
\affiliation{Fachbereich Physik, Freie Universität Berlin, 14195 Berlin, Germany}
\author{Florian Faaber}
\affiliation{Fachbereich Physik and Halle--Berlin--Regensburg Cluster of Excellence CCE, Freie Universit\"at Berlin, Arnimallee 14, 14195 Berlin, Germany}
\author{Paul Wiechers}
\affiliation{Fachbereich Physik and Halle--Berlin--Regensburg Cluster of Excellence CCE, Freie Universit\"at Berlin, Arnimallee 14, 14195 Berlin, Germany}
\author{Caroline Firschke}
\affiliation{Fachbereich Physik and Halle--Berlin--Regensburg Cluster of Excellence CCE, Freie Universit\"at Berlin, Arnimallee 14, 14195 Berlin, Germany}
\author{Tobias Kampfrath}
\affiliation{Fachbereich Physik and Halle--Berlin--Regensburg Cluster of Excellence CCE, Freie Universit\"at Berlin, Arnimallee 14, 14195 Berlin, Germany}
\author{Christian Lotze}
\affiliation{Fachbereich Physik and Halle--Berlin--Regensburg Cluster of Excellence CCE, Freie Universit\"at Berlin, Arnimallee 14, 14195 Berlin, Germany}
\author{Katharina J. Franke}
\affiliation{Fachbereich Physik and Halle--Berlin--Regensburg Cluster of Excellence CCE, Freie Universit\"at Berlin, Arnimallee 14, 14195 Berlin, Germany}

\title{Modified Low-Temperature Scanning Tunneling Microscope for Ultrafast Pump--Probe Spectroscopy}  

\begin{abstract}
Combining scanning tunneling microscopy (STM) with terahertz pulses (THz-STM) enables ultrafast dynamics of phonons, spins, and charge carriers to be probed with atomic-scale spatial and sub-picosecond temporal resolution. However, coupling THz radiation into a low-temperature STM junction presents some technical challenges as it requires optical access for laser illumination while maintaining junction stability. Instead of designing a completely new STM head with customized optical access and optics within the ultra-high vacuum chamber, we modified a Besocke "Beetle" style STM with minimal changes of the cryogenic radiation shields. Despite the given geometrical conditions, we achieved efficient coupling of THz and optical pulses to the STM junction while maintaining temperatures below 6$\,$K. We designed the optical setup on a separate laser table, starting with femtosecond laser pulses that are used to generate THz pulses in a LiNbO$_3$ crystal. We show the obtained THz pulse shapes and determine the effective time resolution using cross-correlation measurements. We further demonstrate the stable performance of the modified STM by measuring THz-induced tunneling currents from monolayer molybdenum disulfide (MoS$_2$) grown on Au(111). The resulting current maps resolve atomic-scale contrast revealing a point defect. The measurements confirm efficient THz coupling and stable low-temperature STM operation with atomic resolution.

\textsuperscript{§}These authors contributed equally to this work.

\end{abstract}
\maketitle

\newpage
\section{Introduction}
Using ultra-short light pulses for nanoscale imaging and spectroscopy with ultrafast time resolution requires confining these pulses beyond the Abbe diffraction limit \cite{abbe_beitrage_1873}. This confinement can be achieved using nanoscale antennas, such as a scanning tunneling microscopy (STM) tip, which convert propagating far-field radiation into strongly localized near fields and, conversely, couple localized near fields to propagating far-field radiation, with the spatial extent mainly determined by the antenna geometry rather than the optical wavelength \cite{wessel_surface-enhanced_1985,stockle_nanoscale_2000,gimzewski1988photon,Berndt1991}. Field-enhanced scanning probe methods, in particular the combination of STM with ultrafast optical or THz pump--probe techniques, exploit this principle and have emerged as powerful approaches for achieving simultaneous sub-millielectronvolt energy, sub-nanometer spatial, and sub-picosecond temporal resolution \cite{grafstrom_photoassisted_2002,yarotski_ultrafast_2002,terada_real-space_2010,yoshida_nanoscale_2012,dolocan_two-color_2011,wu_two-photon-induced_2010,cocker_ultrafast_2013,cocker_tracking_2016,muller_phase-resolved_2020,garg_attosecond_2020,abdo_variable_2021,chen_single-molecule_2023,allerbeck_efficient_2023,li_real-space_2024}. 
Although advantages of near-field excitation for ultrafast STM measurements are obvious, its implementation was only realized several decades after the invention of the STM. The major challenge in establishing ultrafast light-driven STM has been to couple the light pulse into the junction without compromising the junction stability. Therefore, early approaches in fact avoided direct optical excitation. 
For instance, junction-mixing STM uses femtosecond laser pulses that trigger photoconductive switches to generate picosecond voltage pulses that are delivered to the tunnel junction via striplines patterned on the sample \cite{nunes_picosecond_1993,steeves_nanometer-scale_1998,khusnatdinov_ultrafast_2000}. With this approach a temporal resolution of $\approx$ 10\,ps has been achieved, with its limit given by the bandwidth of the microstrip transmission lines. In this scheme, the laser pulses served as "external" voltage triggers, which did not compromise the imaging capabilities of the STM.

In contrast, approaches where the STM tip itself acts as a nanoscale antenna allow the localized near field to directly drive and probe the tunneling current on ultrafast timescales. However, these realizations are often affected by thermal expansions and instabilities in the junction \cite{grafstrom_photoassisted_2002,dolocan_two-color_2011}. Shaken-pulse-pair excitation sequences of optical pulses were used to minimize the thermal effects in all-optical pump--probe schemes of time-delayed femtosecond laser pulse pairs highlighting the potential for probing charge-carrier dynamics at the atomic scale \cite{terada_real-space_2010,dolocan_two-color_2011,yoshida_nanoscale_2012}.

Another breakthrough that minimized thermal effects was achieved by Cocker et al.\cite{cocker_ultrafast_2013} by the invention of terahertz scanning tunneling microscopy (THz-STM). Coupling free-space, single-cycle THz pulses into an STM junction enables sub-picosecond temporal resolution while maintaining atomic-scale spatial resolution. Due to the small THz photon energy, thermal expansion of the tip is minimized, which is particularly important for maintaining junction stability at cryogenic temperatures \cite{tachizaki_progress_2021}.

In brief, the incident THz pulses act as an ultrafast voltage transient ($V_\mathrm{THz}$) across the tunnel junction. The total instantaneous voltage ($V(t)$) is then the sum of the DC bias voltage $V_\mathrm{b}$ and time-varying voltage due to the THz pulse $V_\mathrm{THz}(t)$. This total $V(t)$ drives a time-varying current $I(t)$ = $I_\mathrm{b}$ + $i_\mathrm{THz}(t)$. Due to the non-linearity of the STM junction, the THz-induced current can have a non-zero time integral $\int i_\mathrm{THz}(t) dt \neq 0$, resulting in a rectified current that is detectable in the time-averaged tunnel current. Since we measure the THz-induced tunneling current, which arises from the nonlinear response of the tunnel junction, the spatial resolution is defined by the tunneling area, resulting in the same spatial resolution as conventional STM. 
We note that optical nonlinearities can also be used instead of solely relying on the electrical nonlinearities in a THz-excitation--detection scheme \cite{siday_all-optical_2024}.

The broad applicability of THz-STM is evident from the recent works: THz-field driven tunneling currents have been used to probe the relaxation dynamics of optically excited carriers \cite{cocker_ultrafast_2013, yoshida_subcycle_2019,yoshida_terahertz_2021}, while the ultrafast temporal resolution enabled 100-fs snapshot images of the orbital structure of a single molecule \cite{cocker_tracking_2016}, and of ultrafast coherent molecular motions \cite{cocker_ultrafast_2013,peller_sub-cycle_2020,wang_atomic-scale_2022}. Further findings include the excitation and detection of low-energy collective excitations \cite{sheng_launching_2022}, charge-density-wave dynamics \cite{sheng_terahertz_2024,lopez_ultrafast_2025}, coherent phonon dynamics \cite{roelcke_ultrafast_2024,rai_influence_2025}, charge state manipulation in atomic scale defects \cite{allerbeck_ultrafast_2025}, and the generation of THz-field driven luminescence \cite{kimura_terahertz-field-driven_2021, kimura_ultrafast_2025}. 

A key prerequisite for THz-STM is the efficient coupling of single-cycle, phase-stable THz pulses to the STM junction, while maintaining stable operation at cryogenic temperatures, typically at liquid-nitrogen or liquid-helium temperature \cite{cocker_tracking_2016,yoshioka_real-space_2016,jelic_ultrafast_2017,yoshioka_tailoring_2018,yoshida_subcycle_2019,peller_sub-cycle_2020,muller_phase-resolved_2020,ammerman_lightwave-driven_2021,wang_atomic-scale_2022,allerbeck_efficient_2023}. Given the novelty of this technique, most THz-STM setups are based on customized STM heads and incorporate either adjustable or fixed optical elements that allow the THz pulses to be focused on the STM junction without breaking the vacuum \cite{cocker_ultrafast_2013,yoshioka_real-space_2016,yoshida_subcycle_2019,abdo_variable_2021,zhang_development_2024,azazoglu_variable-temperature_2024,tachizaki_time_2025}.

In this work, we present a home-built THz-STM based on a modified low-temperature ($\sim$4$\mathrm{K}$) Besocke "Beetle" style STM \cite{besocke_easily_1987,frohn_coarse_1989} from CreaTec Fischer and Co. Instead of implementing extensive modifications to the STM head, we modify the nitrogen and helium radiation shields to enable optical access while maintaining operating temperatures of $\sim$6$\mathrm{K}$. We demonstrate efficient coupling of THz pulses with variable repetition rate (1-40$\,$MHz) as well as optical pulses (1030 $\mathrm{nm}$ and 515 $\mathrm{nm}$) into the STM junction. 

In the following, we describe the modified STM setup and the generation and coupling of the THz and optical pulses into the STM junction. We first demonstrate the temporal resolution of the setup by measuring the cross-correlation on an atomically clean Au(111) surface. We then characterize the spatial resolution using atomic scale defects in monolayer molybdenum disulfide (MoS$_2$) grown on Au(111).

\section{Modifications enabling focusing of THz pulses onto the STM junction}

\begin{figure*}
	\centering
		\includegraphics[width=\textwidth]{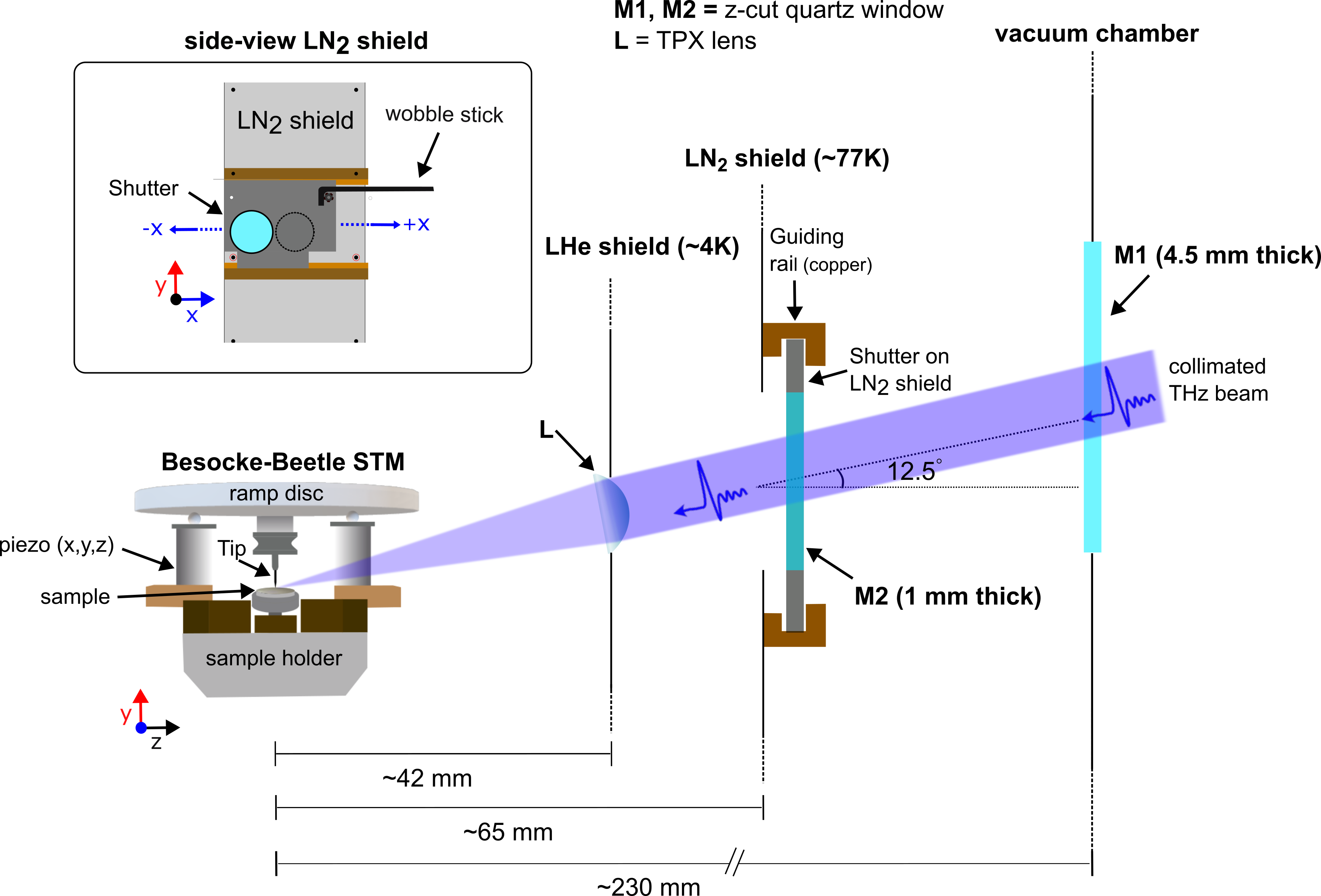}
	\caption{\textbf{Schematic of the THz-STM.} Cross-sectional overview illustrating the path of the THz/optical pulses from outside of the UHV chamber into the STM junction. A collimated THz/optical beam enters the UHV chamber (from right to left) through the quartz window (M1). Thereafter, it passes through the liquid-nitrogen (LN$_2$) and liquid-helium (LHe) radiation shields, respectively. A movable shutter is integrated into the LN$_2$ shield to enable or block the THz/optical beam. Inset (top left): sketch of the shutter mounted on the LN$_2$ shield. Image is not to scale.}
	\label{fig:STM_sketch}
\end{figure*}

Coupling THz or optical pulses into an STM junction while maintaining stable cryogenic temperatures under ultrahigh vacuum (UHV) conditions is often technically challenging. Here we describe a minimally invasive modification to a Besocke "Beetle" style STM, in which optical access to the tunnel junction is achieved while preserving the intrinsic performance and stability of the STM. 

Figure \ref{fig:STM_sketch} shows a schematic of the modified STM setup. In the Besocke "Beetle" style configuration, the central piezo carries the tip and provides the scanning motion, while three surrounding piezo tubes support a
segmented circular ramp for coarse approach. The geometry of the
piezo--tube assembly determines the direction from which optical
access to the junction is possible, and the height of the ramp sets
the incidence angle of the incoming beam. Two radiation shields connected to the liquid-helium and
liquid-nitrogen bath cryostats, respectively, block direct optical access to the
STM junction. To enable optical access, apertures were milled into
both shields, each with a diameter of 40$\,$mm. The positions of these holes were chosen such that the incident collimated beam strikes the junction at an angle of $\approx$ 12.5$^\circ$ with respect to the sample surface. 

In the hole of the LHe shield, a 25.4$\,$mm TPX: polymethylpentene lens ($n = 1.45$) with a focal length of 42$\,$mm is mounted. The lens is permanently attached to the shield at such an angle that the incident beam is parallel to its optical axis. This lens focuses the THz beam onto the tip--sample junction when the tip is roughly positioned at the center of the sample. The diffraction-limited spot size of the focused THz beam is approximately 1$\,$mm at 1$\,$THz. This size corresponds to a focus depth of about 2$\,$mm, providing sufficient range in the $x-z$ plane for positioning the tip at different locations on the sample while maintaining efficient coupling of the THz pulses. Likewise, the width of the THz spot in the $x-y$ plane is sufficiently large for covering the movement of the STM tip in $z$ direction, which is limited to $\pm$300$\,$$\mu$m. 

\begin{figure}
	\centering
		\includegraphics{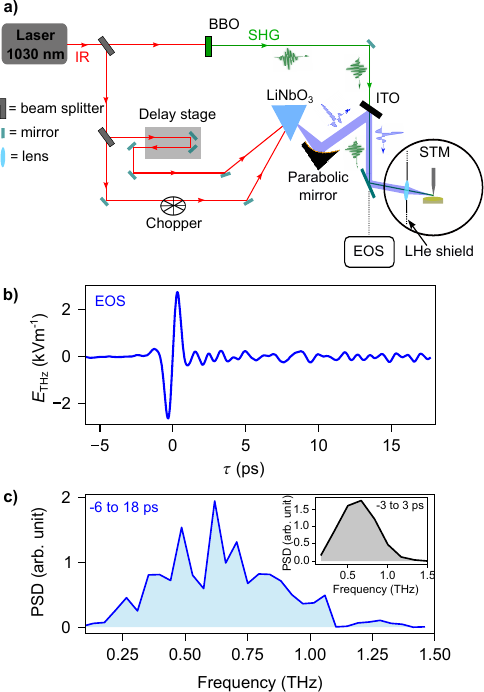}
	\caption{\textbf{Schematic of the optical setup for THz generation and far-field characterization.} \textbf{(a)} Simplified sketch of the THz generation setup. A nitrogen purge box encloses the THz beam path to minimize the absorption by atmospheric water (humidity), marked with dotted box. \textbf{(b)} EOS shows the single-cycle THz pulse. \textbf{(c)} FFT performed on the pulse shown in \textbf{b}, taken in the temporal window of -6$\,$ps to 18$\,$ps. FFT performed in a smaller window, centered around 0$\,$ps is shown in the inset.}
	\label{fig:setup_EOS}
\end{figure}

The alignment of the incident collimated THz beam with a diameter of $\approx$76.2$\,$mm, which is defined by the off-axis parabolic mirror after THz pulse generation (for details on the optical table, see Fig.\ref{fig:setup_EOS}a), provides an additional degree of freedom for covering a sufficiently large range of focus positions. By horizontally or vertically shifting the incident beam, the focal point at the sample can be moved, allowing THz measurements to be performed at different positions on the sample surface without changing the optical alignment inside the cryostat. 

To reduce liquid-helium consumption, there is no permanent hole in the LN$_2$ shield. Instead, a z-cut quartz window (M2) is mounted on a movable shutter. This shutter can be moved horizontally on a rail with a wobble stick (see the inset of Fig. \ref{fig:STM_sketch}) to align the window with the milled hole in the nitrogen shield. When optical access is not required, the shutter can be moved to close the window to reduce the liquid-helium consumption and reduce the base temperature.

A z-cut quartz window of 63$\,$mm diameter and thickness of 4.5$\,$mm with a refractive index $n = 2.1$ at 1\,THz is mounted on the vacuum chamber (marked with M1 in Fig.\ref{fig:STM_sketch}) to allow the THz beam to enter. 

The modifications described here are applicable to most variants of conventional STM designs which have visual access from a UHV view port, while keeping the mechanical structure of the microscope unchanged. Thus, the noise performance and low temperature baseline (without radiation) of the STM should remain unaffected, while adding substantial benefit of ultrafast time resolution to the system.

\section{Generation of the single-cycle phase stable THz pulses}
To utilize the THz pulses as voltage transients with sub-picosecond temporal resolution, phase-stable, single-cycle THz pulses are required. Several THz generation schemes have been employed in previous THz-STM studies and instrumentation reports, including optical rectification of femtosecond laser pulses in lithium niobate (LiNbO$_3$) \cite{meyer_single-cycle_2020,wulf_analysis_2021}, in spintronic emitters \cite{seifert_efficient_2016,vogel_efficient_2022,muller_phase-resolved_2020}, or in photoconductive emitters \cite{chen_single-molecule_2023,azazoglu_variable-temperature_2024}. 
High-field THz pulses are well suited for driving strong-field
processes, such as the excitation of molecular orbitals to probe
molecular dynamics \cite{cocker_tracking_2016}. In contrast, low-field THz pulses are often required to investigate dynamics
governed by low-energy excitations such as phonons\cite{roelcke_ultrafast_2024,rai_influence_2025} or spins. Because low-field pulses
generate correspondingly small THz-induced tunneling currents,
high repetition rates are essential to achieve a sufficient signal-to-noise ratio \cite{sheng_terahertz_2024,wang_atomic-scale_2022}.

For our setup, we target operation at small THz-field amplitudes and variable high repetition rates. 
A simplified schematic of the THz generation setup is shown in Fig.\ref{fig:setup_EOS}a. The fundamental near-infrared (NIR) laser source (Amplitude Satsuma HP3) delivers pulses with a duration of <350 $\mathrm{fs}$ and a central wavelength of 1030 $\mathrm{nm}$ at a tunable repetition rate ranging from 1 to 40 $\mathrm{MHz}$. This primary beam is first split into two paths (labeled as P$_1$ and P$_2$ in Fig.\ref{fig:setup_EOS}a), one for optical excitation and pulse characterization and another one for THz generation in a pump--probe scheme. 

The first part of the split beam serves as the optical pump (P$_1$) and is directed to the STM junction for optical-pump--THz probe spectroscopy. It can also be used for far-field THz pulse characterization by electro-optic sampling (EOS) outside the UHV chamber and for junction-based measurements, such as photoemission sampling (PES) \cite{yoshida_subcycle_2019} and photoexcitation dynamics \cite{yoshida_terahertz_2021,lopez_ultrafast_2025,jelic_femtosecond_2026}. For PES, green 515 $\mathrm{nm}$ pulses are generated from the fundamental 1030 $\mathrm{nm}$ beam by second-harmonic generation in a beta-barium borate (BBO) crystal.

The second part of the NIR beam (P$_2$) is dedicated for THz generation. It is split once more to establish a THz-THz pump-probe configuration. In this scheme, one beam propagates across a motorized delay stage before recombining with the non-delayed beam. The pulse fronts of both beams are tilted by an optical grating to match the phase velocity of the THz pulse and group velocity of the pump IR beam inside the LiNbO$_3$ crystal \cite{hebling_generation_2008}. An LiNbO$_3$ crystal is ideal for THz generation from variable femtosecond-pulse repetition rates because optical rectification in LiNbO$_3$ exhibits a robust linear relationship between the incident
IR pulse energy and the emitted THz field \cite{abdo_variable_2021}. The THz pulses are directed via mirrors towards the STM. The last part of this setup is housed in a small box that can be purged with nitrogen gas to minimize THz absorption by water molecules in the air (see dotted box in Fig. \ref{fig:setup_EOS}a).

An indium tin oxide (ITO) mirror, which is transparent to visible and IR light, but reflects THz radiation, is used to combine the optical and THz beams to the same optical path. A rotatable silver mirror (labeled M$_R$ in Fig.\ref{fig:setup_EOS}a) is used to direct both beams towards the STM or to a stage for EOS. Finally, a pair of wire-grid polarizers precisely controls the amplitude and polarization of the THz pulses before they are guided into the UHV chamber and coupled into the STM junction.  

To understand the response of the tunnel junction to the incident THz pulses and the subsequent dynamics, it is important to characterize the pulse shapes in as much detail as possible \cite{yoshida_subcycle_2019,ammerman_lightwave-driven_2021,azazoglu_coupling_2026}. A rough approximation to the pulse in the junction can be found in the far-field. The far-field waveform of the generated THz pulses can be characterized using EOS. By simply rotating a silver mirror positioned in front of the UHV window of the STM chamber, both the IR and THz beams are redirected toward the EOS detection setup outside of the UHV chamber.  

The collinear NIR and THz beams propagate through a gallium phosphide (GaP) crystal. Inside this nonlinear crystal, the transient THz electric field distorts the lattice structure, thereby modulating the polarization of the co-propagating NIR sampling beam. The resulting elliptically polarized beam is separated into its $s$- and $p$-polarization components using a quarter-wave plate and a Wollaston prism. A balanced photodetector measures the intensity difference between these two components, which can be directly converted into the far-field THz amplitude \cite{wu_freespace_1995,jepsen_detection_1996}. 

The waveform of a typical THz pulse, acquired via EOS is displayed in Fig.\ref{fig:setup_EOS}b. The measured waveform reveals the single-cycle nature of the THz pulse, followed by small amplitude oscillations. These originate from absorption and re-emission of THz radiation from atmospheric water molecules \cite{cheville_far-infrared_1999}. The fast Fourier transform (FFT) of the pulse in a temporal window of -6$\,$ps to 18$\,$ps is shown in Fig.\ref{fig:setup_EOS}c. The effect of THz absorption by water molecules is reflected in the FFT with a clear characteristic dip at $\sim$ 0.55 THz. The FFT in a narrow temporal window from -3$\,$ps to 3$\,$ps shows a broad-band peak around a central frequency of $\sim$ 0.6$\,$THz (inset of Fig.\ref{fig:setup_EOS}c). This bandwidth is ideal for investigating low-energy excitations such as phonons or spins.

\section{Performance of the THz-STM}
\subsection{Thermal stability}
Coupling the generated THz beam into the STM junction, while maintaining thermal stability is crucial. Thermal expansion within the junction has traditionally been a challenge in adapting visible or infrared laser pulses for time-resolved STM measurements \cite{grafstrom_photoassisted_2002,dolocan_two-color_2011}. In contrast, THz pulses induce minimal heating due to their low photon energy, making them highly advantageous for probing ultrafast dynamics in an STM junction at cryogenic temperatures.

\begin{figure}
	\centering
		\includegraphics{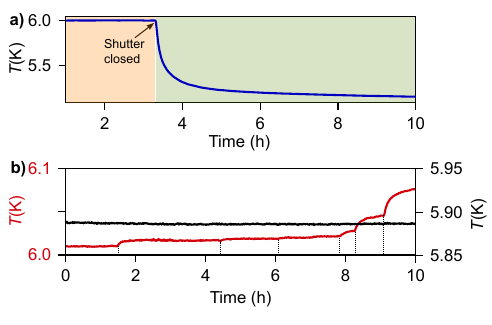}
	\caption{\textbf{Temperature stability.} \textbf{(a)} Temperature curve showing a constant temperature of $\sim$ 6$\,$K (orange shaded region) when the shutter is open and THz pulses at minimum field (i.e. $\Delta\theta_{wgp} = 90^{\circ}$) are impinging on the STM junction. The green shaded region shows the temperature decrease after the shutter is closed. \textbf{(b)} The red curve shows the temperature as the impinging THz field is increased by opening the wire grid polarizer angle ($\Delta\theta_{wgp}$). Each step in the temperature curve (marked by vertical dotted lines) corresponds to an decrease in $\Delta\theta_{wgp}$ by $10^{\circ}$ in the range of $0.067E_{max}$ ($\Delta\theta_{wgp} = 75^{\circ}$) to $0.93E_{max}$ ($\Delta\theta_{wgp} = 15^{\circ}$). The black curve shows the temperature stability over a period of 10$\,$h at a constant THz field.}
	\label{fig:Temperature}
\end{figure}

We first check the thermal stability by measuring the temperature at the base plate of the STM head. When the shutter in the liquid-nitrogen shield is closed, a base temperature of $\sim$5.2\,K is reached. With an open shutter, the temperature rises to $\sim$6\,K, which recovers to the base temperature within a few hours after the shutter has been closed again (Fig.\,\ref{fig:Temperature}a). 
Because the system remains highly stable with only a small increase in temperature by $\sim$0.8$\,$K in the open-window configuration, we conclude that the integration of the optical window in the liquid-nitrogen shield and the lens in the liquid-helium shield does not compromise the overall cryogenic performance. 

The next crucial step is to maintain a stable temperature in the presence of THz radiation. Figure\,\ref{fig:Temperature}b shows the temperature response to varying electric field amplitudes, where the THz field amplitude was tuned by adjusting the relative angle ($\Delta\theta_\mathrm{wgp}$) between the two wire-grid polarizers (with the latter being fixed to assure a well-defined polarization of the field along the tip axis). Upon increasing the THz amplitude by opening the wire grid in $10^{\circ}$ increments across six steps (marked with vertical dotted lines in Fig.\ref{fig:Temperature}b, corresponding to $0.067E_{max}$ ($\Delta\theta_{wgp} = 75^{\circ}$) to $0.93E_{max}$ ($\Delta\theta_{wgp} = 15^{\circ}$)), the temperature rises slightly but stabilizes after a few tens of minutes (Fig.\ref{fig:Temperature}b). This demonstrates that the THz-STM can operate stably under low-temperature conditions.

\subsection{THz pulse shape at STM junction: photoemission sampling}
\begin{figure}
	\centering
		\includegraphics{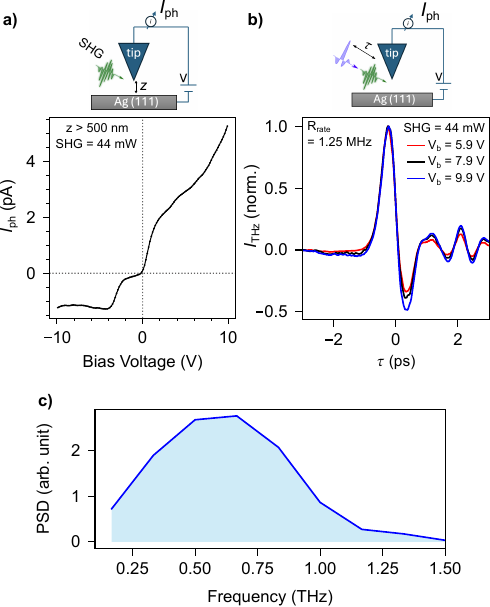}
	\caption{\textbf{THz pulse shape in STM junction via photoemission sampling.} \textbf{(a)} Second harmonic generation (SHG)-induced photocurrent as a function of applied DC bias voltage in the STM junction consisting of a Ag tip and Ag(111) substrate with tip retracted by more than 500$\,$nm from the tunneling regime. \textbf{(b)} Three normalized THz waveform measurements, normalized to their respective maximum positive peaks, acquired at DC bias voltages of 5.9$\,$V (red), 7.9$\,$V (black), 9.9$\,$V (blue). For the waveform sampling, the time delay ($\tau$) between the SHG and THz pulse is swept. \textbf{(c)} FFT of the THz waveform (blue curve in \textbf{b}) in the temporal window of -3$\,$ps to 3$\,$ps.}
	\label{fig:PES}
\end{figure}

Having characterized the THz waveform incident on the tip by EOS and demonstrated that the STM remains thermally stable under illumination with the THz radiation, we need to characterize the actual pulse shape that drives the tunneling current in the STM junction. The free-space THz pulse shape can be altered by the antenna effect of the tip through a transfer function that depends on the THz coupling geometry, as well as the shape and size of the tip \cite{muller_phase-resolved_2020}. The frequency-dependent transfer function of the tip depends on the combined resistive, capacitive, and inductive coupling, which may lead to spectral filtering and phase shifts. Because the temporal profile of the electric field in the junction directly determines the achievable time resolution, it is essential to characterize the specific THz waveform \cite{yoshida_subcycle_2019,ammerman_algorithm_2022,li_real-space_2024}. 

The waveform of this THz near-field transient within the STM junction can be characterized using photoemission sampling (PES) \cite{yoshida_subcycle_2019}. For this measurement, the STM tip coated with silver (prepared by controlled indentations into a Ag(111) substrate) is retracted by more than 500$\,$nm above the Ag(111) surface. A 350-fs green optical pulse (515 nm, generated as described above) is focused onto the tip apex, inducing a photocurrent via a two-photon absorption process. This photocurrent strongly depends on the DC bias voltage ($V_\mathrm{b}$) applied between the tip and the sample surface; the corresponding $I$-$V$ curve, obtained by sweeping the DC sample bias, is shown in Fig.\ref{fig:PES}a. Notably, at positive DC bias voltages, the current is positive and increases with voltage, indicating that electron emission originates from the tip apex. Similarly, at negative DC bias voltages, the current is negative, indicating electron emission from the substrate. 

To measure the THz pulse shape at the tip, the DC bias voltage is fixed where the static photocurrent $I$-$V$ response is approximately linear. Under such conditions, the transient THz voltage modulates the photocurrent linearly, allowing the measured current modulation to be directly representing the THz pulse shape. Figure \ref{fig:PES}b displays optical--THz delay traces recorded at different DC bias voltages. In the photon--driven regime described by Keldysh theory \cite{keldysh_ionization_1965}, the optical pump generates a short photocurrent burst, much shorter than the THz transient, which therefore samples the instantaneous THz field at the STM junction. Scanning the relative delay between the optical and THz pulse maps out the THz waveform at the tip, such that the shape of the resulting curve represents the THz near-field waveform. However, care must be taken when interpreting the measured waveform, as deviations from a linear photocurrent response with DC bias or a THz transient voltage that drives the junction into the nonlinear regime can distort the waveform. This is evident from the waveforms measured at different DC bias voltages (see Fig.\ref{fig:PES}b), where the relative amplitudes of the positive and negative half-cycles vary with the applied DC bias voltage. Nevertheless, the near-field transient amplified by the STM tip exhibits a robust single-cycle profile with a central frequency of $\sim$ 0.6 THz. The corresponding FFT of the pulse shape in the temporal window -3$\,$ps to 3$\,$ps is shown in Fig.\ref{fig:PES}c.

\subsection{Temporal resolution of the THz-STM}

\begin{figure}
	\centering
		\includegraphics{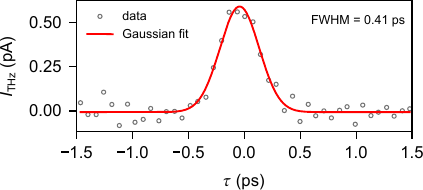}
	\caption{THz-induced tunneling current measured in a THz cross-correlation scheme on a clean Au(111) surface. The parameters used for recording the time-delayed data: $V_\mathrm{b}$=2.8$\,$V, $I_\mathrm{t}$=300$\,$pA, and $R_\mathrm{rate}$=5$\,$MHz. The feedback was switched off during the measurement, and the time-delayed beam was chopped at 867$\,$Hz. A Gaussian fit was performed to get the full width at half maximum (FWHM) of the cross-correlation pulse (red curve).}
	\label{fig:Temporal}
\end{figure}

Though the pulse shape already suggests a sub-ps time resolution (by estimating the width of the first half cycle of the pulse), other factors, such as the non-linearities of the junction and the effective peak width of the pulse that actually induces the THz current, determine the lower limit of the time resolution. 
As a good system for probing the time resolution in a THz-pump--probe scheme, we chose the strong non-linearity at the onset of a field-emission resonance on a Au(111) surface. 
We then perform cross-correlation measurements of two time-delayed THz pulses, where the first THz pulse induces a field-emission current, which is modulated by a much smaller second pulse (which would not be sufficient to induce a field-emission current by itself) \cite{peller_quantitative_2021,li_real-space_2024}.  
Figure~\ref{fig:Temporal} shows the THz-induced current as a function of time delay between two pulses ($E_1 =$ 367  V/m, and $E_2 =$ 43.4 V/m). A Gaussian fit to the cross-correlation trace yields a full width at half maximum (FWHM) of 410\,fs. The THz-induced maximum tunneling current by the probe pulse at $\tau$ = 0$\,$ps reaches the value of $\approx$0.55$\,$pA, corresponding to $\approx$0.45$\,$electrons/pulse. This sub-picosecond current modulation demonstrates the potential of THz-driven spectroscopy for probing ultrafast dynamics with sub-picosecond temporal resolution.

\subsection{Spatial resolution of the THz-STM}
\begin{figure}
	\centering
		\includegraphics{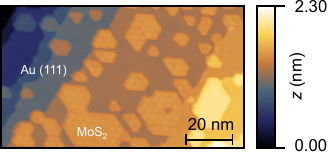}
	\caption{Large area topography of MoS$_2$ nanoislands grown on Au(111) surface. Tunneling parameters for recording the topography are $V_\mathrm{b}$ = 1.0\,V, and $I$ = 35\,pA.} 
	\label{fig:Large_scan}
\end{figure}

To demonstrate the spatial resolution of our modified THz-STM setup, we investigated monolayer islands of molybdenum disulfide (2H-MoS$_2$) grown on Au(111). Molybdenum was deposited onto a clean Au(111) substrate in an H$_2$S environment at a pressure of 10$^{-5}$$\,$mbar, followed by annealing to 550$^\circ$C for 1\,hr \cite{gronborg_synthesis_2015,krane_moire_2018}. The monolayer-MoS$_2$ islands exhibit a moiré structure with a periodicity of 3.33\,nm due to the lattice mismatch between the Au substrate and the S layer of the MoS$_2$ (Fig.\ref{fig:Large_scan}) \cite{gronborg_synthesis_2015,sorensen_structure_2014}. 

\begin{figure}
	\centering
		\includegraphics{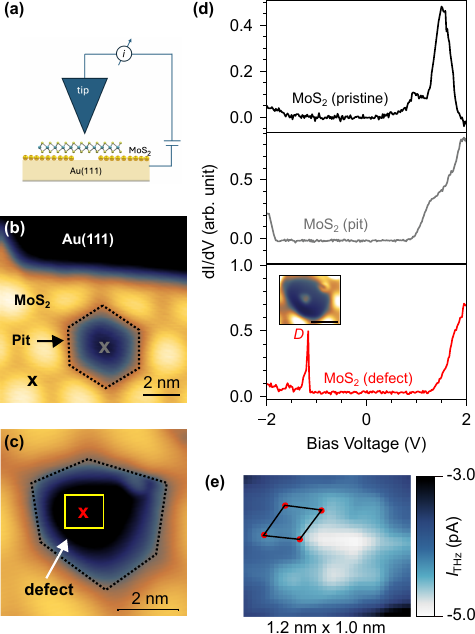}
	\caption{\textbf{Spatial resolution of THz-STM.} \textbf{(a)} Sketch of monolayer MoS$_2$ on Au vacancy island. \textbf{(b)} Topography of MoS$_2$ island containing a pit ($V_\mathrm{b}$ = 1.0\,V, and $I$ = 100\,pA). \textbf{(c)} Pit with a defect ($V_\mathrm{b}$ = 1.0\,V, and $I$ = 50\,pA). \textbf{(d)} d$I$/d$V_\mathrm{b}$ spectra recorded on the pristine MoS$_2$ (black), on the pit (gray) and on a pit containing a defect (red). Measurement positions for the spectra are marked in \textbf{b} and \textbf{c} with respective colors. Inset in the bottom panel shows the constant height map at $V_\mathrm{b}$ = -1.15\,V (scale bar is 3$\,$nm). \textbf{(e)} THz induced current ($I_\mathrm{THz}$) map recorded on the defect present in the pit. Area on which the map is recorded is marked with yellow rectangle in \textbf{c}. The diamond (in black) corresponds to the unit cell of MoS$_2$.}
	\label{fig:Spatial_res}
\end{figure}

On the MoS$_2$ islands, we often observe dark-appearing areas several nanometers in size at moderate bias voltages. This indicates a larger band gap than the regular islands. A more detailed electronic-structure analysis led to the conclusion that these areas are quasi-freestanding nanopatches of MoS$_2$ due to Au vacancies beneath the MoS$_2$ layer \cite{krane_electronic_2016} (see sketch in Fig.\ref{fig:Spatial_res}a). 
 Owing to their appearance, they were referred as “pits” \cite{krane_electronic_2016}. Two of them are shown in Fig.\ref{fig:Spatial_res}b and c. Solely from the topographies at low bias voltage, the two pits appear very similar. However, their electronic structure differs significantly. The d$I$/d$V_\mathrm{b}$ spectra measured on the two pits together with the pristine region are shown in Fig.\ref{fig:Spatial_res}d. The spectrum in black is measured on the pristine MoS$_2$, where the onsets of the conduction and valence bands are consistent with previous reports \cite{sorensen_structure_2014,bruix_single-layer_2016,krane_electronic_2016}. The gray spectrum is measured on the pit shown in Fig.\ref{fig:Spatial_res}b and shows a much wider band gap, consistent with its quasi-freestanding nature \cite{krane_electronic_2016}. Interestingly, d$I$/d$V_\mathrm{b}$ on the second pit (Fig.\ref{fig:Spatial_res}c) shows a sharp positive ion resonance at $V_\mathrm{b}$ = -1.17$\,$V, indicating the presence of a defect state. 

From the STM topography and the d$I$/d$V_\mathrm{b}$ alone, it is not trivial to determine the nature of this defect. Moreover, the topography acquired at the positive-ion resonance (PIR) energy of the defect does not reveal any distinct structural features that can be associated with the typical top- or bottom-layer sulfur vacancies.\cite{mitterreiter_role_2021, trishin_electronic_2023} (see inset of Fig.\ref{fig:Spatial_res}d). We therefore use THz-STM to further investigate the nature of this defect. Fig.\ref{fig:Spatial_res}e shows a constant-height THz-induced current ($I_\mathrm{THz}$) map, acquired over the region marked by the yellow rectangle in Fig.\ref{fig:Spatial_res}c. For this measurement, the DC bias voltage was set within the electronic gap of the pit ($V_\mathrm{b}$ = -0.9\,V, where the topography would look dark), and the feedback was switched off. The incident THz field was set to $E=\mathrm114 $ V/m by using the wire grid polarizer angle $\Delta\theta_{wgp} = 50^{\circ}$ ($0.41E_{max}$). The $I_\mathrm{THz}$ map shows a pronounced spatial variation in the THz current over the defect, with a contrast showing atomic resolution. However, the observed structure deviates from the expected MoS$_2$ unit cell (black diamond in Fig.\ref{fig:Spatial_res}e), suggesting local distortion of the  atomic lattice. A pronounced bright feature is also observed at the defect center. Such distortion might be a consequence of defect in the second atomic layer, such as a Mo vacancy. Such an assignment is consistent with previous reports of Mo vacancies exhibiting a similar electronic structure (see Fig.\ref{fig:Spatial_res}d) \cite{trainer_visualization_2022}. 
A more comprehensive characterization of the defect would require measurements of its spatial and energetic behavior, which are beyond the scope of the present report. Importantly, this atomic-scale imaging capability of THz-STM can be used for visualizing ultrafast atomic-scale lattice motion directly in real space \cite{jelic_terahertz_2025}.

The resolution of the atomic lattice in the constant height THz current map clearly demonstrates the high junction stability and atomic spatial resolution of our THz-STM. This stability is essential for resolving the spatial variation of the THz induced current without the contributions from feedback driven tip-height modulation. 

\section{Conclusion}
We have implemented rather simple modifications to a Besocke "Beetle" style STM, limited to providing optical access to the tunnel junction by integrating windows and lenses into the radiation shields. Despite of the stringent geometrical constraints imposed on the beam path, this design enabled efficient coupling of optical and THz pulses into the junction. Importantly, neither the modifications themselves nor the laser illumination compromised the STM's stability. 

We chose to generate the THz pulses by rectification in LiNbO$_3$ and showed a time resolution of $\sim$400\,fs. 
However, we emphasize that the STM setup is compatible with other methods of pulse generation, such as spintronic emitters \cite{muller_phase-resolved_2020} or even shorter pulses \cite{garg_attosecond_2020}, which could improve the time resolution. 

A key consideration in the choice of the laser and THz pulse generation scheme was the possibility to operate at high repetition rates of up to 40\,MHz. This ensures that pulses that induce less than one electron per pulse are still easily detectable via a lock-in technique. The ability to work with small THz-induced currents is particularly important for the investigation of low-energy phenomena.

\section{Acknowledgment}
We thank Sebastian Loth for valuable comments and hints during the setup and sharing his experience with his THz-STM. We acknowledge financial support by the Deutsche Forschungsgemeinschaft (DFG, German Research Foundation) through Project No. 328545488 (CRC 227, Project No. B05).

\bibliographystyle{model1-num-names.bst}

\bibliography{references}

\end{document}